\documentclass[amsmath,amssymb,floatfix,nopreprintnumbers,nofootinbib,prd,onecolumn,preprint,superscriptaddress]{revtex4}
\usepackage{hyperref}
\usepackage{graphicx}
\usepackage[usenames,dvipsnames,svgnames]{xcolor}
\usepackage{amsmath}
\usepackage{amssymb}

\hypersetup{
  colorlinks,
  linktoc=page,
  linkcolor=blue,
  citecolor=ForestGreen
}

\newcommand{\beq}{\begin{equation}}
\newcommand{\eeq}{\end{equation}}
\newcommand{\tr}{{\rm Tr}}

\begin{document}

\title{Review of strongly-coupled composite dark matter models \\ and lattice simulations}

\author{Graham D.~Kribs}\email{kribs@uoregon.edu}
\affiliation{Department of Physics, University of Oregon,
Eugene, OR, 97403 USA}
\author{Ethan T.~Neil}\email{ethan.neil@colorado.edu}
\affiliation{Department of Physics, University of Colorado, 
Boulder, CO, 80309, USA}
\affiliation{RIKEN-BNL Research Center, 
Brookhaven National Laboratory, Upton, NY 11973, USA}

\begin{abstract}

We review models of new physics in which dark matter arises as a 
composite bound state from a confining strongly-coupled non-Abelian
gauge theory.  We discuss several qualitatively distinct classes 
of composite candidates, including dark mesons, dark baryons, 
and dark glueballs.  We highlight some of the promising strategies 
for direct detection, especially through dark moments,
using the symmetries and properties of the composite 
description to identify the operators 
that dominate the interactions of dark matter 
with matter, as well as dark matter self-interactions.  
We briefly discuss the implications of these theories at
colliders, especially the (potentially novel) phenomenology 
of dark mesons in various regimes of the models. 
Throughout the review, we highlight the use of lattice 
calculations in the study of these strongly-coupled theories, 
to obtain precise quantitative predictions and new insights 
into the dynamics.  

\end{abstract}

\maketitle

\tableofcontents

\section{Introduction}
	\label{sec:DM_intro}
	
Dark matter (see recent reviews of particle dark matter\cite{Hooper:2009zm,Lisanti:2016jxe}) is an essential ingredient in current models of the evolution of our universe, with strong evidence for its existence found from galactic to cosmological scales.  Aside from its gravitational interactions, the precise nature of particle dark matter has remained elusive, but two key properties have now been strongly established.  Dark matter must be stable over more than the lifetime of the universe.  Dark matter must essentially also both be electrically neutral (from cosmology) and effectively neutral with respect to the standard model (from direct detection experiments).

From a theoretical standpoint, these two crucial properties of dark matter motivate certain properties of any particle dark matter model.  The required stability suggests the existence of some appropriate symmetry in the dark sector which prevents or greatly suppresses dark matter decay.  As for interaction with the standard model, although it is possible that dark matter has only gravitational interactions, the cosmic coincidence problem (namely, that the relic abundance of dark matter and ordinary matter are of the same order of magnitude) motivates an early-universe connection which is stronger than gravity.

There are two instructive examples to be found in the standard model: the proton and the neutron.  The proton, although it is charged, is known to have an extremely long lifetime.  Its stability is a consequence of the accidental baryon-number symmetry of the standard model - any operator which would cause the proton to decay is non-renormalizable and thus suppressed by the scale of new physics.  The neutron is made of charged fundamental components, the quarks, but itself has zero net electric charge.  Its interactions with the photon are strongly suppressed at low energies, but at high temperatures in the early universe its constituent quarks interact at full strength with electromagnetism.

These observations lead naturally to the idea of strongly-coupled composite dark matter.  A dark sector with non-Abelian gauge interactions and fundamental constituents charged under the standard model (or some other mediator) can give rise to a stable, neutral dark matter candidate which can interact more strongly in the early universe to explain the cosmic coincidence problem.  Such a sector containing a composite dark matter candidate can arise quite naturally from extensions of the standard model in which the Higgs boson is composite \cite{Contino:2010rs,Andersen:2011yj,Bellazzini:2014yua}, or could simply be their own ``dark sector''.  Composite dark matter can naturally have a substantial self-interaction cross section, which has been invoked as one possible explanation of observed galactic structure anomalies \cite{Spergel:1999mh,Vogelsberger:2012ku,Rocha:2012jg,Peter:2012jh,Zavala:2012us}.  The dynamics of composite dark matter can lead to interesting and unexpected phenomena, for example, the first realization of the idea of asymmetric dark matter was in the context of technicolor theories \cite{Nussinov:1985xr}.

From the perspective of the low energy effective theory below the 
compositeness scale, whatever forms the dark matter can of course
be treated as an elementary particle with a series of higher 
dimensional operators, allowed by all symmetries of both the 
dark sector and the standard model fields.  In particular,
for direct detection the non-relativistic effective theory
has been written in a general form\cite{Fan:2010gt,Fitzpatrick:2012ix,Cirigliano:2012pq,Hill:2014yka,Hill:2014yxa}.  This gives all possible
interactions in a well-defined expansion in momentum.

Of course, although the basic structure of this effective theory is dictated only by low-energy
symmetries, some operators may be suppressed by features of the ultraviolet theory.
The advantage of working with specific ``ultraviolet completions''
is to determine which operators \emph{are} or \emph{are not} generated
for a given theory, and then use appropriate techniques -- including
lattice simulations -- to determine the coefficients of the operators.  
This knowledge can be used to relate bounds or predictions from 
one class of experiments, such as direct detection, to (or from) 
another class, such as collider experiments.

Lattice calculations are an important tool for understanding composite dark matter models.  Due to their strongly-coupled nature, perturbative methods are not generally applicable in such a model.  Lattice gauge theory allows for precise, quantitative study of many interesting quantities, including the full spectrum of composite states and matrix elements important for direct and indirect detection, relic abundance, self-interactions, and collider production.  Lattice realizations also require working with ultraviolet-complete models, which depend on a handful of fundamental parameters as opposed to the numerous couplings that can appear in a low-energy effective description.  As an integral part of this review, we will highlight existing lattice calculations and opportunities for future lattice studies in the context of various models and experimental searches.

This review is organized as follows: in section 2, we consider various composite dark matter candidates which can arise from strongly-coupled theories, classifying them broadly as ``meson-like", ``baryon-like", and ``glueball-like".  Details about how the correct relic abundance is attained and the possibility of ``dark nuclear" bound states are also discussed.  Section 3 focuses on direct detection of composite dark matter, in particular through suppressed couplings to standard model force carriers.  Section 4 turns to the rich collider phenomenology of composite dark sectors, discussing current bounds and future prospects.  Finally, in section 5 we give an outlook on interesting future directions, both for model builders and for lattice simulations.

\section{Composite dark matter candidates}
	\label{sec:DM_candidates}
	
In this section we review strongly-coupled theories that contain 
composite dark matter candidates.\footnote{We do not consider
theories with weakly-coupled bound states, such as ``dark atoms.''
\cite{Kaplan:2009de,Khlopov:2015nrq}}  
There are numerous reasons to
seriously consider strongly-coupled composite dark matter,
and we highlight the major motivations in the following.  Not all models
necessarily contain all or most of the following characteristics;
instead we are simply providing a broad brush of the interesting 
consequences.  Other reviews which focus on different aspects 
of strongly-coupled composite dark matter can be found in the 
literature \cite{Cline:2013zca,Lewis:2014boa,Antipin:2015xia}.

\subsection{Motivation}

\textbf{Dark Matter Stability}.  One of the principle attractions of 
composite dark matter candidates is that stability can be an automatic
consequence of the accidental global flavor symmetries of the 
underlying theory.  Once above the compositeness scale, 
the (gauge) symmetries of the dark constituents may 
restrict the dimension of the leading operators 
$d$ that violate the global flavor symmetries 
to be sufficiently high that even with maximal violation by 
Planck-suppressed operators (e.g.,  $\mathcal{O}_d/M_{Pl}^{d-4}$), 
the dark matter stability is far longer than the age of the universe. 
This is, after all, the reason that proton stability is well understood
in the standard model, despite possible baryon number violation 
by Planck-suppressed operators.  

\textbf{Naturalness}.  Just like QCD, once a non-Abelian theory 
confines, a new scale appears through dimensional transmutation,
the dark confinement scale $\Lambda_d$.  This scale is technically 
natural, allowing for an effective theory description of the 
infrared theory -- the dark mesons and baryons -- through operators 
suppressed by this scale.  

\textbf{Dark Matter Neutrality}.  In theories where the 
constituents transform under (part of) the standard model, 
confinement can lead to color, weak, and charge-neutral
dark hadrons that provide candidates for neutral dark matter.
This typically imposes constraints on the parameters of the 
underlying theory.\footnote{This is not unlike analogous constraints
on theories of elementary dark matter candidates, in which the lightest dark parity-odd particle is
required to be neutral under the standard model.}

\textbf{Suppressed interactions}.  The effective theory  
below the confinement scale can be expressed in terms of
higher dimensional operators involving (pairs of) dark matter
fields with standard model fields, suppressed by powers of
the dark confinement scale.  This can provide a beautiful mechanism
to suppress dark matter scattering off nuclei, below the
tight experimental bounds that exist from direct detection experiments.  
The coefficients 
of the dark moments can in principle be computed from the underlying 
ultraviolet theory.  This is a major motivation for lattice
simulations that can provide superior estimates of the coefficients
of the dark moments over naive dimensional analysis power counting. 

\textbf{Self-interactions}.  Strongly-coupled theories naturally
have strong self-interactions among the mesons and baryons.
If the scales are arranged appropriately, these self-interactions
may be responsible for addressing the observed galactic structure
anomalies\cite{Spergel:1999mh,Vogelsberger:2012ku,Rocha:2012jg,Peter:2012jh,Zavala:2012us}.
This has provided
a strong motivation for recent consideration of strongly-coupled
self-interacting dark matter
\cite{Carlson:1992fn,Cline:2013zca,Boddy:2014yra,Hochberg:2014dra,Boddy:2014qxa,Soni:2016gzf}.

\textbf{New observables}.  The rich spectrum of dark hadrons   
that appear after the dark non-Abelian theory confines
provide a plethora of experimental targets.  This includes
novel detection strategies such as inelastic scattering to 
excited states \cite{ArkaniHamed:2008qn,Alves:2010dd,Kumar:2011iy}, 
dark absorption lines \cite{Profumo:2006im,Kribs:2009fy},
effects on the CMB and $N_{\rm eff}$ 
\cite{Garcia:2015loa,Buen-Abad:2015ova}
and a host of collider phenomenology consequences (to be discussed below).  We note that although not focused on dark matter models, spectroscopy of SU$(N)$
gauge theories in the large-$N$ limit has been studied extensively on the lattice \cite{Panero:2012qx,DeGrand:2013nna,Lucini:2014bwa, Cordon:2014sda}.

\subsection{Meson dark matter I:  Pion-like}

There are three broad classes of composite dark matter made from 
mesons of a confining, strongly-coupled non-Abelian group:
pion-like ($m_q \ll \Lambda_d$), quarkonia-like
($m_q \gg \Lambda_d$), and an intermediate regime
($m_q \sim \Lambda_d$) or mixed regime ($m_{q_1} < \Lambda_d < m_{q_2}$). 
Meson stability relies on accidental dark flavor 
(or ``species'' \cite{Kilic:2009mi}) symmetries.  The dark flavor
symmetries could be continuous or discrete, such as $G$-parity
\cite{Bai:2010qg}. 

Several models of pion-like dark matter have been proposed 
\cite{Ryttov:2008xe,Hambye:2009fg,Bai:2010qg,Lewis:2011zb,Buckley:2012ky,Frigerio:2012uc,Bhattacharya:2013kma,Hochberg:2014kqa,Hietanen:2014xca,Carmona:2015haa,Hochberg:2015vrg}.
One reason for their popularity is familiarity from QCD, 
and specifically, utilizing chiral effective theory techniques
to characterize the mass spectrum and pion interactions.  
In the following, we describe only a selection of models that
have been proposed.

Weakly interacting stable pions was proposed in Ref.\cite{Bai:2010qg}.
In this theory, stability is ensured through $G$-parity,
that is a modified charge conjugation operation allowed
when using real representations of the SM gauge group.
Dark fermions transform in vector-like
representations of an $SU(N)_d \times SU(2)_L$,
where the reality of $SU(2)$ representations permits
$G$-parity to be preserved in the Lagrangian.  Pions transform 
as $\Pi^{(J M)} \stackrel{G}{\rightarrow} (-1)^G \Pi^{(J M)}$,
and thus the lightest $G$-odd pion, $\Pi^{(1 \, 0)}$,
is a dark matter candidate.  At the level of the 
chiral Lagrangian, $\Pi^{(1 0)}$ does not decay through 
the usual axial anomaly (unlike the standard QCD case, 
$\pi^0 \rightarrow \gamma\gamma$), due to the vanishing of
the relevant isospin trace.  

A different proposal to use weakly interacting pions was
proposed in \cite{Buckley:2012ky}.  Dark fermions transform
in vector-like representations of $SU(2)_d \times U(1)_Y$.
The use of $SU(2)$ for the new confining group 
(called ``ectocolor'' \cite{Buckley:2012ky}) has the feature that there
are five pseudo-Goldstone bosons -- three are the usual pions, 
while the remaining two can be identified as ``baryon-like''
dark matter in the theory.  Of course no baryons result once 
$SU(2)_d$ confines, however, a global $U(1)_X$ symmetry
can be imposed to ensure the baryon-like pions are 
stable with respect to the low energy effective theory.
Once again, chiral Lagrangian techniques can be used to calculate
the leading scattering cross sections and decay rates of
the pions.  That the dark matter baryon-like pions are
close in mass to the pions that do not carry a conserved global
$U(1)_X$ number leads to a novel, nontrivial freezeout process
effectively driven by co-annihilation among the light pion
species \cite{Griest:1990kh,Buckley:2012ky}. 
This leads to weak scale masses and pion decay
constants that can be probed by collider experiments.

More recently, a strongly-coupled theory with pions as dark matter
was proposed (``SIMP'' for strongly interacting massive particle) 
\cite{Hochberg:2014kqa,Hochberg:2015vrg}.
Again, chiral Lagrangian techniques allow the estimation
of the leading pion interactions.  Here, the novel feature
that occurs for a wide class of non-Abelian theories 
($SU(N_c)$ or $SO(N_c)$ with $N_f \ge 3$; $Sp(N_c)$ with $N_f \ge 2$)
in the pion-like limit is that there is a 5-point interaction arising 
from the Wess-Zumino-Witten action enabling a $3 \rightarrow 2$ 
annihilation process.
Unlike the earlier proposals, all of the dark fermions are 
neutral under the SM, and so a new mediator is necessary to 
connect the thermal bath of the standard model with this dark sector.
In \cite{Hochberg:2015vrg}, a global $U(1)$ flavor symmetry of the theory
is gauged (and broken explicitly) leading to a massive ``dark photon''
with assumed kinetic mixing with hypercharge.  Under suitable conditions,
the $3 \rightarrow 2$ process is active and leads to the freeze out
of dark matter.  Intriguingly, the relevant scales of the 
new non-Abelian dark sector that lead to the correct dark matter 
relic abundance is very similar to QCD, $m_\pi \sim 300$~MeV
with $f_\pi \sim \mathrm{few} \times m_\pi$.  Constraints on
the dark photon mass and kinetic mixing that enable this 
mechanism were presented in \cite{Hochberg:2015vrg}.  Principally,
the pions cannot decay into dark photons. This can be guaranteed 
if all the pions transform non-trivially under part of the 
unbroken flavor symmetry.  And, near degeneracy of the dark quarks
is required, so that at least five different pions participate 
in the $3 \rightarrow 2$ process via the WZW term.

\subsection{Meson dark matter II:  Quarkonium-like}

In the regime where there is at least one heavy dark fermion
with mass $m_q > \Lambda_d$, heavy quark effective theory
can be applied, and qualitative differences from the
pion-like theories result.

One example of this class of model is ``composite inelastic dark matter''
\cite{Alves:2009nf,Lisanti:2009am,Alves:2010dd}, where dark matter
is a meson made from one light and one heavy quark.  In this theory,
the hyperfine interactions split the ground state by a small
energy that can be relevant to dark matter direct detection.
The possibility that dark matter may have its dominant interaction
with nuclei through an inelastic scattering process is 
well known \cite{TuckerSmith:2001hy,TuckerSmith:2004jv,Chang:2008gd}. 
Strongly-coupled composite theories provide a natural home for
small inelastic transitions, and this can lead to a rich 
spectroscopy.  

In the regime where all of the dark fermions are heavier than
the confinement scale leads to another class of composite 
dark matter generally known as ``quirky
\footnote{The name ``quirky'' came from 
a fascinating class of theories that postulate new dark 
fermions that transform under part of the standard model, and also 
transform under a new non-Abelian group that confines 
at a scale far below the mass of the fermions.\cite{Kang:2008ea}} 
 dark matter'' \cite{Kribs:2009fy}. 
The dark fermions were taken to be in a chiral representation 
of the electroweak group, using the Higgs mechanism to give them mass.
For the specific theory of SU(2), 
it was known \cite{Peskin:1980gc,Preskill:1980mz}
that confinement aligned the vacuum towards an electroweak 
preserving minimum, and thus not substantially affecting electroweak
symmetry breaking.  In addition, with the bound states containing
exactly two heavy dark fermions, a perturbative non-relativistic 
treatment of the composite dark matter mesons is possible.  
This allowed an estimate of the excited meson masses, 
as well as the coefficients of the effective operators
leading to quirky dark matter scattering with nuclei.
Quirkonium production and decay were considered in 
\cite{Harnik:2011mv,Fok:2011yc}. 

One drawback to dark matter composed of dark mesons is the
potential difficulty in maintaining the exactness
of the global flavor quantum number that ensures that
the dark matter is (sufficiently) stable.  For example, 
already at dimension-5 there can be operators that violate 
global flavor symmetries
\begin{eqnarray}
\frac{1}{\Lambda} \overline{\Psi} \Psi H^\dagger H \quad , \quad
\frac{1}{\Lambda} \overline{\Psi} \sigma^{\mu\nu} \Psi B_{\mu\nu} \, . 
\label{eq:dim-5-ops}
\end{eqnarray}
Here $\overline{\Psi}\Psi$ is a fermion bilinear that transforms
nontrivially under the global flavor symmetry that protects against
meson decay in the effective theory.  
Even with $\Lambda = M_{\rm Pl}$, these operators with order one coefficients
lead to dark meson lifetimes much shorter than the age of the 
universe.  Of course there are ways to suppress these interactions,
but it requires additional model-building at higher scales.

\subsection{Baryon-like dark matter}

One of the principle reasons to consider baryon-like candidates
for dark matter is robust stability, i.e., safety 
from higher dimensional interactions that lead to decay on
timescales short compared with the age of the universe, 
c.f. Eq.~(\ref{eq:dim-5-ops}).
For theories with fermion constituents, $SU(N_c)$ 
with $N_c \ge 3$, higher dimensional operators are at least
dimension-6 or higher.  In these theories, dark matter is automatically
sufficiently stable, and no further ultraviolet model-building is
needed.  This is a superior property of composite baryonic dark matter.

Early work on technicolor theories demonstrated the potential 
of technibaryons as a dark matter candidate
\cite{Nussinov:1985xr,Chivukula:1989qb,Barr:1990ca}. 
In these theories,
dark fermions transformed under a chiral representation of
$SU(2)_L \times U(1)_Y$, and so after confinement, lead 
to dynamical electroweak symmetry breaking.  The technibaryons
carried an accidental global quantum number, technibaryon number,
that suggested the lightest technibaryon is a natural dark matter 
candidate.  Early investigation in these theories revealed an 
elegant mechanism to obtain the correct cosmological abundance of 
dark matter through a technibaryon number asymmetry
\cite{Barr:1990ca,Barr:1991qn,Kaplan:1991ah}. 
These investigations continued into aspects of direct detection
\cite{Chivukula:1992pn,Bagnasco:1993st,Pospelov:2000bq}.  
More recent investigations into technibaryon dark matter and 
other related candidates can be found
\cite{Dietrich:2006cm,Gudnason:2006yj,Nardi:2008ix,Foadi:2008qv,Ryttov:2008xe,Khlopov:2008ty,Sannino:2009za,Mardon:2009gw,Lewis:2011zb,Cline:2013zca,Hietanen:2013fya,Brod:2014loa,Hietanen:2014xca}.

With the discovery of the Higgs boson in 2012, technicolor
theories, at least as originally formulated, are under siege.
(Theories that lead to a Higgs-like boson, such as composite Higgs theories, 
remain interesting, but also have substantial constraints from
the LHC.)  This leaves open the possibility of theories containing
approximately vector-like fermions transforming under (part of) the 
standard model, using the strong dynamics to set mass scales
as well as to provide a viable electroweak-neutral composite dark 
matter candidate.

The LSD Collaboration has investigated both fermion and scalar 
baryonic candidates for dark matter from confining SU(3) and SU(4) 
dark color theories
\cite{Appelquist:2013ms,Appelquist:2015yfa,Appelquist:2015zfa}. 
In both cases, dark fermions were
assumed to transform under (nearly) vector-like representations
of the electroweak group, leading to negligible corrections to
electroweak precision observables.  In the dark SU(3) case, 
the lightest baryon was found to be a neutron-like fermionic
dark baryon with a significant magnetic moment due to the 
electrically charged dark fermion constituents.
The spectrum and the leading interactions with nuclei
were determined using lattice simulations.  
Given existing bounds from Xenon experiments, the lower bound 
on the mass of this fermionic dark baryon was found to be
$10$~TeV \cite{Appelquist:2013ms}.  This comparatively large mass
arises due to the relatively low mass dimension of the magnetic
moment interaction (dimension-5).  

In \cite{Appelquist:2015yfa,Appelquist:2015zfa}, the LSD 
Collaboration proposed and investigated scalar dark baryons from an 
SU(4) dark confining interaction, called ``Stealth Dark Matter''.  
When combined with a dark custodial SU(2) symmetry (that leads to 
equal masses for the lightest $q = \pm 1/2$ electrically charged 
dark fermion constituents),
stealth dark matter was found to be remarkably safe from 
direct detection experiments due to the high dimension of the leading 
interaction -- electromagnetic polarizability -- of the 
scalar baryon with the standard model.  Dark baryons as light as
$300$~GeV were possible for an order one pion-to-vector mass ratio.
These relatively precise estimates were possible by performing
lattice simulations to determine the hadron mass spectrum
as well as the coefficients of the dominant operators
leading to direct detection.  
Given that the polarizability-induced spin-independent direct detection 
cross section scales as $Z^{8/3}$, heavier element experiments 
(including xenon and tungsten) clearly have better sensitivity than 
lighter elements (such as germanium and argon).  In addition, this theory 
contains a rich spectrum of mesons somewhat below the mass
of the dark baryon that are ripe for exploration at the LHC.

\subsection{Dark glueballs}

Any non-Abelian gauge sector is also expected to contain a number of glueball bound states which have no valence fermion content.  In theories such as QCD where the colored fermions are light compared to the confinement scale, these glueballs are broad resonances which decay readily into lighter mesons and baryons, and often mix with neutral mesons as well.  These properties make QCD glueballs rather difficult to isolate, and no conclusive experimental observation has been reported to date \cite{Crede:2008vw}.

However, if all fermions in a dark non-Abelian sector are very heavy compared to the confinement scale $\Lambda_d$, then the lightest particles in the spectrum will be the glueballs, with masses of order $\Lambda_d$.  Like baryonic states, the lightest glueballs are stabilized by accidental symmetry, since as color-singlet bound states their creation operators are dimension 4, of the form $\tr (G_{\mu \nu} G^{\mu \nu})$, or dimension 6 of the form $\tr(G_{\mu \nu}^3)$.  The leading operators in the Lagrangian which can mediate glueball decay are then of the form \cite{Faraggi:2000pv}
\beq
\mathcal{L} \supset \frac{c_H}{M^2} H^\dagger H \tr (G_{\mu \nu} G^{\mu \nu}) + \frac{c_F}{M^4} \tr (G_{\mu \nu} G^{\mu \nu}) \tr (F_{\mu \nu} F^{\mu \nu}), 
\eeq
where $F_{\mu \nu}$ is the field-strength tensor for one of the standard model gauge fields, and $M$ is the scale of new physics, e.g. some heavy fermions which carry both standard model and hidden sector charge.   These operators mediate glueball decays which scale as $\Gamma \sim c_H \Lambda_d^9 / M_H^4 M^4$ and $\Gamma \sim c_F \Lambda_d^9 / M^8$ respectively, so that the glueballs can be stabilized for a wide range of $\Lambda_d$, as long as $M$ is sufficiently large: for example taking $M = M_{\rm GUT} = 10^{16}$ GeV, the glueballs will be stable on the lifetime of the universe if $\Lambda_d \lesssim 10^{5}$ GeV \cite{Faraggi:2000pv}.  A detailed effective field theory description of glueball decay processes has been studied in \cite{Juknevich:2009ji,Juknevich:2009gg}. 

Hidden sector glueballs are thus a natural dark matter candidate \cite{Okun:1980kw,Okun:1980mu}, with a number of interesting properties.  Hidden sectors of this type can fit nicely into larger models of new physics, e.g. as part of the MSSM with anomaly-mediated supersymmetry breaking \cite{Feng:2011ik,Boddy:2014yra,Boddy:2014qxa} - in which case the dark matter consists of both glueballs and glueballinos, with abundance depending on parameter choices.  Grand unification can also lead to hidden non-Abelian sectors (see e.g. \cite{Kakushadze:1996jm,Kakushadze:1997ne} for a partial classification); GUT and string-theory motivated studies of glueball dark matter are undertaken in \cite{Faraggi:2000pv, Yamanaka:2014pva}.

Collider bounds on glueball dark matter are relatively weak compared to other composite candidates, because unlike the mesonic and baryonic cases, no light charged states exist in the spectrum if all of the hidden-sector fermions are heavy.  The absence of strong constraints from LEP makes it much easier to construct viable glueball dark matter models with a dark matter mass below 100 GeV.  In particular, the strong self-interactions of the glueballs can provide an explanation of galactic structure anomalies \cite{Spergel:1999mh,Vogelsberger:2012ku,Rocha:2012jg,Peter:2012jh,Zavala:2012us} for glueball dark matter masses in the MeV to GeV range, depending on the specific choice of hidden gauge group \cite{Boddy:2014yra, Boddy:2014qxa,Soni:2016gzf}.

Although there have been no lattice calculations to date specifically focused on glueball dark matter, there are a number of more general results on the glueball spectrum and selected matrix elements \cite{Morningstar:1999rf,Lucini:2004my,Chen:2005mg,Loan:2006gm,Lucini:2010nv, Lucini:2014paa}.  Since any hypercolor-charged particles are taken to be heavy compared to the confinement scale, from the perspective of a lattice practitioner the theory of interest is ``pure-gauge" Yang-Mills; this is an attractive theory to study, because in the absence of fermions large-scale studies can be undertaken with relatively modest computational resources.

\begin{figure}[t]
\centering
\label{fig:glue-3}
\begin{minipage}[b]{0.45\textwidth}
	\includegraphics[width=\textwidth]{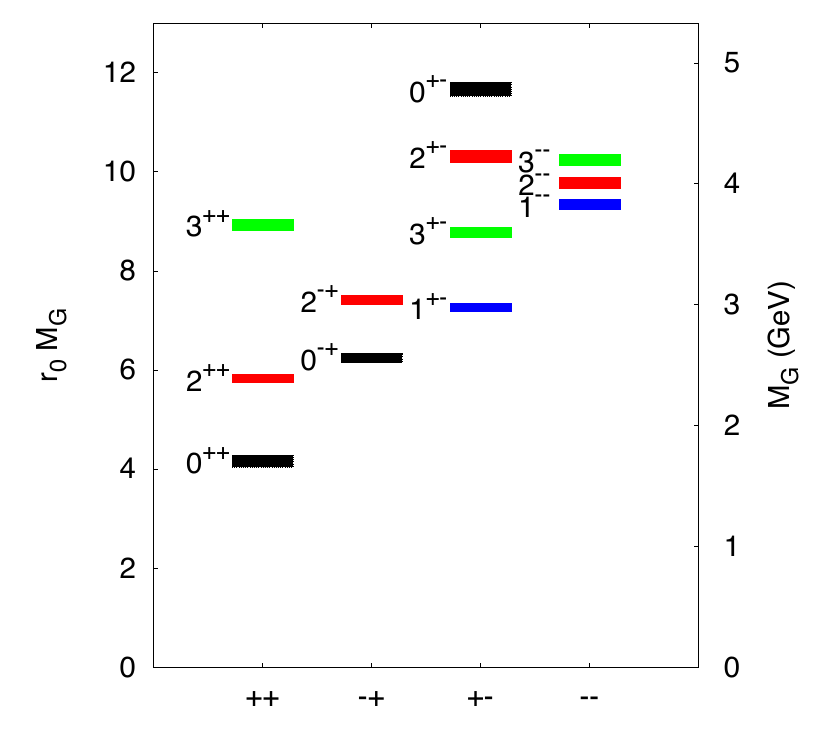}
\end{minipage}
\hfill
\begin{minipage}[b]{0.54\textwidth}
	\includegraphics[width=\textwidth]{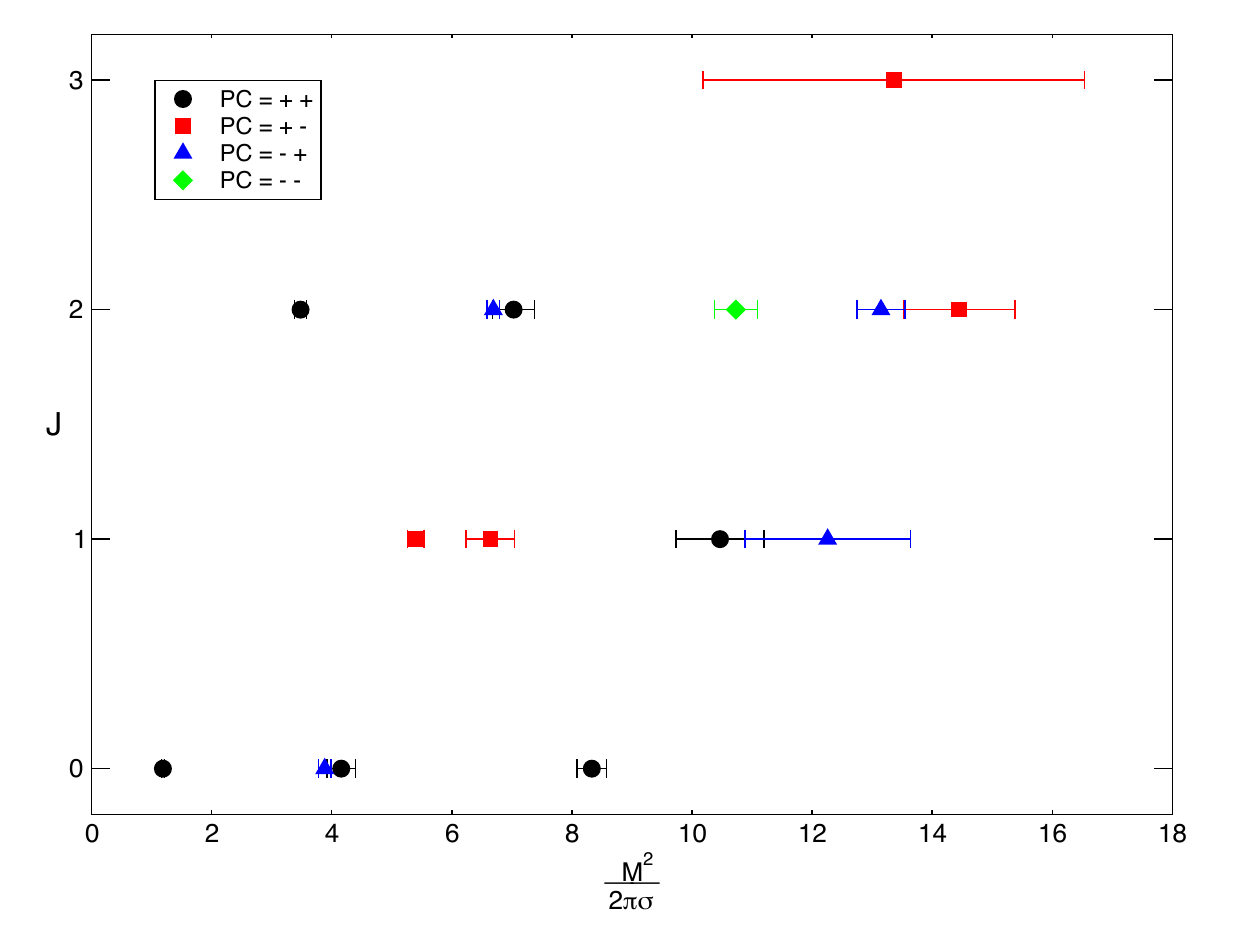}
\end{minipage}
\caption{Glueball spectrum obtained from lattice simulations for SU$(3)$ pure gauge theory \cite{Chen:2005mg} (left) and for SU$(N)$ pure gauge theory extrapolated to the large-$N$ limit \cite{Lucini:2010nv} (right).}
\end{figure}

\subsection{Abundance}

To obtain a cosmologically significant thermal relic abundance
of dark matter, the standard approach is to allow the strongly-coupled
composites to annihilate into other, light, unstable dark sector 
states that decay back into standard model particles.  
The annihilation cross sections are generically large,
due to strong coupling.  If the annihilation rate saturates
the unitarity bound\cite{Griest:1989wd,Blum:2014dca},
the dark matter mass is on the order of $100$~TeV\@.  
If the dark sector contains many states close in mass,
the annihilation rates can be substantially modified, 
and the mass of the dark matter can be smaller,
closer to the TeV scale.\cite{Buckley:2012ky},  
A more recent exception 
to the usual $2 \rightarrow 2$ strongly-coupled annihilation rate 
is when a $3 \rightarrow 2$ 
process dominates the thermal freezeout of dark pions.  For this
process to be operable, both a
new interaction with standard model (e.g.\ dark photons kinetically
mixed with hypercharge) is required as well as degeneracy of the 
lightest flavors of dark fermions to ensure the effective 
Wess-Zumino-Witten 5-point pion interaction is unsuppressed.

An asymmetric abundance of strongly-coupled dark matter 
was considered long ago in the context of technibaryon 
dark matter \cite{Barr:1990ca,Barr:1991qn,Kaplan:1991ah}.  
That technibaryons transformed under a chiral representation 
of the electroweak group implies that electroweak sphalerons
preserve the linear combination $U(1)_{B - L - D}$,
opening up the possibility that an asymmetric abundance
of dark baryons is automatically generated following
baryogenesis or darkogenesis 
\cite{Barr:1991qn,Kribs:2009fy,Shelton:2010ta,Davoudiasl:2010am,Buckley:2010ui}.
Intriguingly, the residual abundance of dark baryons is 
$\rho \sim m_B n_B$ where the number density is 
proportional to $\exp[-m_{\rm B}/T_{\rm sph}]$, 
where $T_{\rm sph}$ is the temperature
at which sphaleron interactions shut off. 
If the baryon and dark baryon number densities are comparable,
the would-be overabundance of dark matter (from $m_B \gg m_{\rm nucleon}$)
is compensated by the Boltzmann suppression.  Very roughly,
$m_{\rm baryon} \sim 1$-$2$~TeV
is the natural mass scale that matches the 
cosmological abundance of dark matter \cite{Barr:1990ca}.  A similar mass scale
can be arrived at through dynamics coupling the QCD scale to the dark confinement scale
\cite{Bai:2013xga}.

In theories with vector-like and electroweak symmetry breaking
masses, such as stealth dark matter
\cite{Appelquist:2015yfa,Appelquist:2015zfa},
it was anticipated that an asymmetric abundance was still
possible, but with further suppression by the amount of
electroweak symmetry breaking in the dark sector.

\subsection{Dark Nuclei and Dark Nucleosynthesis}

Strongly-coupled hadrons from a dark sector may combine to form stable
composites of the hadrons themselves: dark nuclei.  Certainly
the standard model provides a clear proof of principle that
such nuclei exist and provide an order one fraction of the
energy density of matter.  In the presence of a light mediator,
``darkleosynthesis'' was shown to be possible and 
efficient for asymmetric composite dark matter from a 
confining non-Abelian theory\cite{Krnjaic:2014xza}. 
A quantitative exploration of dark nuclei 
was performed for a dark SU(2) with two
flavors.\cite{Detmold:2014qqa,Detmold:2014kba}
Lattice simulations of this model 
demonstrated stable nuclear states are possible with the lowest 
lying states being bound states of the pion and vector mesons 
and their baryonic partners.  This suggests the possibility of 
analogues of nuclei should be considered in any strongly interacting 
composite model.  

In \cite{Detmold:2014qqa,Detmold:2014kba}, it was shown that
for both symmetric and asymmetric origins of dark hadrons,
the early universe cosmology can be substantially altered by
dark nucleosynthesis, perhaps having most dark nucleons processed
into dark nuclei.  Importantly, new signals of indirect detection 
of asymmetric dark matter are possible as dark hadrons combine
into dark nuclei, emitting photons [either directly or as a result
of $U(1)_Y$ kinetically mixed with a dark U(1)].  Other exotic
phenomena may also occur, such as the ejection of asymmetric 
dark nuclei from stars, thereby suppressing the accumulation 
of asymmetric dark matter in these objects.  It remains to be 
seen how generalizable these results are to different numbers of 
color, flavors, and dark fermion mass spectrum.

Finally, an intriguing possibility is that strongly-coupled
dark baryons could form very large dark nuclei, forming an
extended semi-uniform object \cite{Hardy:2014mqa,Hardy:2015boa}.
This is counter to the usual intiution from the standard model,
where only light elements form from big-bang nucleosynthesis. 
It was found that dark nuclei with large dark nucleon number,
$A \gtrsim 10^8$ may be synthesized. \cite{Hardy:2014mqa,Hardy:2015boa}. 
This qualitatively changes direct detection and capture rates in 
astrophysical objects.

\section{Direct detection of composite dark matter}
	\label{sec:DM_direct}
	
Direct detection experiments have long since ruled out dark matter with spin-independent elastic scattering through an electroweak-strength tree-level coupling to the $Z$ boson.  However, the standard model gauge bosons are not ruled out as mediators for standard model-dark matter scattering if their couplings are suppressed.  In composite dark matter models, the occurrence of this suppression naturally occurs through form factors $F(q^2)$, where $q^\mu$ is the momentum transfer through the interaction vertex.  A form factor can be thought of as a momentum-dependent coupling between a force carrier and a composite particle; at $q^2 = 0$ the form factor is equal to the net charge of the composite, while at large $q^2$ it probes the internal constituents.

The interaction of dark matter with ordinary matter mediated by a standard model force carrier can give a distinctive scaling of the interaction cross-section with the choice of nuclear target, depending on the exact nature of the interaction.  A strongly distinguishable case would be photon exchange coupling to the nuclear magnetic moment, which can give a direct detection cross section varying by several orders of magnitude among commonly-used target elements.  More complete effective field theory treatments of how a general dark matter particle can interact with nuclear direct-detection targets have been considered in the literature \cite{Fan:2010gt,Fitzpatrick:2012ix,Cirigliano:2012pq,Hill:2014yka,Hill:2014yxa,Dent:2015zpa}.  For very light (sub-GeV) dark matter, which can arise from some composite models, scattering off of electrons can provide stronger direct-detection bounds \cite{Essig:2011nj,Graham:2012su,Essig:2015cda,Hochberg:2015pha,Hochberg:2015fth} 

We will focus in this review on photon and Higgs-mediated direct detection through moments of the dark matter.  $Z$-boson exchange proceeds through dark moments similar to those for photon exchange, but is additionally suppressed by the $Z$ mass.  Gluon-based interactions are an interesting possibility which have been considered in the literature \cite{Chivukula:1992pn,Godbole:2015gma,Bai:2015swa}.

\subsection{Photon interactions}

If the composite dark matter candidate $\chi$ is neutral, but its constituents carry electromagnetic charge, then its coupling to the photon is proportional to the matrix element
\beq
\langle \chi(p') | j_{\rm EM}^\mu | \chi(p) \rangle = F(q^2) q^\mu,
\eeq
where $q_\mu = p_\mu + p'_\mu$, $j_{\rm EM}^\mu$ is the electromagnetic current, and $F(0) = 0$.  In the limit that the momentum transfer $|q|$ is very small compared to the compositeness scale $\Lambda$, which is appropriate for dark matter direct detection, the form factor can be described in terms of effective field theory operators of increasing dimension.  The leading C and P-conserving operators are \cite{Bagnasco:1993st,Pospelov:2000bq} the magnetic moment
\beq
\mathcal{L} \supset \frac{1}{\Lambda} \bar{\chi} \sigma^{\mu \nu} \chi F_{\mu \nu},
\eeq
the charge radius
\beq
\mathcal{L} \supset \frac{1}{\Lambda^2} \bar{\chi} \gamma^\nu  \chi \partial^\mu F_{\mu \nu},\ \ \frac{1}{\Lambda^2} \phi^\dagger \phi v^\nu \partial^\mu F_{\mu \nu},
\eeq
and the electromagnetic polarizability
\beq
\mathcal{L} \supset \frac{1}{\Lambda^3} \bar{\chi} \chi F_{\mu \nu} F^{\mu \nu},\ \ \frac{1}{\Lambda^3} \phi^\dagger \phi F_{\mu \nu} F^{\mu \nu},
\eeq
where $\chi$ represents a fermionic dark matter candidate, and $\phi$ a spin-zero bosonic candidate.  If $\phi$ were a boson with non-zero spin, it would also have a magnetic moment operator.  (Note that the dimensions of the operators are the same for scalar dark matter, due to the non-relativistic normalization of the fields which is most appropriate for treating dark matter direct detection \cite{Hill:2014yxa}.)  Other operators at similar orders in the effective expansion, e.g. an electric dipole moment, can appear if CP violation occurs in the dark sector \cite{Pospelov:2000bq}.  Phenomenological treatments of dark matter with some or all of these effective interactions have been considered in the literature, both independently and in the context of composite models \cite{Bagnasco:1993st,Pospelov:2000bq,Sigurdson:2004zp,Gardner:2008yn,Chang:2010en,Banks:2010eh,Barger:2010gv,DelNobile:2012tx,Weiner:2012cb,Pospelov:2013nea,Ovanesyan:2014fha}. 

Detailed formulas for the interaction cross sections mediated by these interactions are derived in the above references.  We will not reproduce them here, but it is worth observing that the scaling of the cross section with the choice of nuclear target can be dramatically different \cite{Dent:2015zpa,DelNobile:2015tza,DelNobile:2015rmp}, depending on which operator dominates.  The per-nucleon interaction cross section is expected to scale as $\mu^2 (J+1)/J$, $Z^2 / A^2$, and $Z^4 / A^{8/3}$ for the magnetic moment, charge radius, and electromagnetic polarizability operators respectively, where $\mu$ is the nuclear magnetic moment, $J$ is the nuclear spin, and $Z$ and $A$ are the standard proton and atomic mass numbers.  (Note that for a dark matter magnetic moment, the scaling given is for the moment-moment interaction; there is also a magnetic moment-nuclear charge interaction \cite{Banks:2010eh,Appelquist:2013ms}, which scales as $Z^2/A^2$ like the charge radius.)

The value of these prefactors for several nuclear targets currently used in direct detection experiments are tabulated in Table~\ref{tab:EM_scaling_normalized}, scaled so that the value for xenon is set to 1.  Especially dramatic differences are seen for the coupling to the nuclear magnetic moment.  We also note that the electromagnetic polarizability interaction in principle has a very large uncertainty; since the interaction contains two photons, scattering proceeds through a loop diagram, so this interaction may be particularly sensitive to poorly-known nuclear matrix elements involving excited states \cite{Appelquist:2015zfa}.

On the dark matter side, determination of the coefficients of these operators requires a non-perturbative calculation.  We now turn to lattice calculations focused on photon direct-detection operators.

\begin{table}[ph]
\caption{Leading scaling of direct-detection interactions involving photon exchange: dark magnetic moment-nuclear moment (first column), dark magnetic moment-nuclear charge or dark charge radius (second column), and dark electromagnetic polarizability (third column), relative to the given prefactor for xenon.  We average over natural isotopic abundance for each element, taking the nuclear magnetic moment $\mu$ and spin $J$ from the literature \cite{Fuller:1976xx}.}
{\begin{tabular}{c|cccc}
\hline
target &	$\mu^2 (J+1)/J$	& $Z^2/A^2$	& $Z^4/A^{8/3}$ \\
\hline
Xe 	&1			&1		&1		\\
Si	&0.06681		&1.472	&0.2766	\\
Ge	&0.1130		&1.152	&0.6010	\\
Na	&12.68		&1.357	&0.1798	\\
O	&0.003029	&1.482	&0.1323	\\
I	&17.09		&1.033	&1.018	\\
Ca	&0.004658	&1.476	&0.4464	\\
W	&0.009074	&0.9608	&1.442	\\
Ar	&0.			&1.201	&0.2949	\\
C	&0.02518		&1.481	&0.0900	\\
F	&32.07		&1.331	&0.1341	\\
\hline
\end{tabular}
\label{tab:EM_scaling_normalized}
}
\end{table}

Calculation of the magnetic moment and charge radius for a given dark matter candidate can be accomplished through a direct lattice calculation of the form factor $F(Q^2)$ itself.  That is, the three-point correlation function
\beq
C_3^\mu(t,t') = \sum_{\vec{x}, \vec{y}} e^{-i \vec{p}' \cdot \vec{x}} e^{-i (\vec{p}-\vec{p}') \cdot \vec{y}} \langle B^\dagger(0,0) V^\mu(\vec{x}, t) B(\vec{y}, t') \rangle
\eeq
is computed directly, where $B$ is the composite object interpolating operator and $V^\mu$ the electromagnetic current.  Calculating at several values of the discretized momentum transfer and fitting the momentum dependence allows the magnetic moment and charge radius to be determined.

Results on the lattice have been obtained for SU$(2)$ \cite{Hietanen:2013fya} and SU$(3)$ \cite{Appelquist:2013ms} gauge theories.  In the former case, the calculated charge radius for the meson-like dark matter candidate is found to be roughly consistent with its predicted value from vector meson dominance, using the value of the vector-meson mass determined from the lattice as well.  Fairly strong bounds are found from the Xenon100 and LUX experiments, although their model has an additional adjustable parameter $d_b$ which can suppress the charge radius interaction (with $d_b = 0$ corresponding to the restoration of an isospin-like symmetry.)  For the SU$(3)$ study, strong bounds are found particularly from the magnetic moment, restricting the dark matter mass to be larger than roughly 10 TeV from Xenon100 constraints alone. 

Determination of the electromagnetic polarizability can be somewhat more difficult, due to its suppression by large powers of the momentum transfer.  An alternative to direct calculation of the form factor is to apply the background field method \cite{Detmold:2010ts}.  In this approach, a background static electric field $\mathcal{E}$ is applied by use of appropriate boundary conditions in the lattice simulation.  Measuring the ground-state energy of the dark matter candidate as a function of $|\mathcal{E}|$ allows determination of the polarizability from the quadratic Stark shift, e.g.\cite{Appelquist:2015zfa}
\beq
E_X(|\mathcal{E}|) = m_X + \left( 2C_F - \frac{\mu_X^2}{8m_X^3} \right) |\mathcal{E}|^2 + \mathcal{O}(|\mathcal{E}|^4), 
\eeq
where $C_F$ is the polarizability and $\mu_X$ is the magnetic moment of $X$.

This approach has been used so far to study two different theories on the lattice.  The LSD collaboration has calculated the polarizability in SU$(4)$ gauge theory for their ``stealth dark matter" model \cite{Appelquist:2015zfa}.  In units of the SU$(4)$ baryon mass, the polarizability was found to be comparable to that of the neutron in QCD, much larger than naive dimensional analysis would indicate.  The resulting direct-detection cross section in LUX diminishes rapidly with the dark matter mass and falls below the expected cosmic neutrino background, but an interesting window for direct detection remains below 1 TeV or so.  The polarizability has also been studied in an SU$(2)$ gauge theory \cite{Drach:2015epq} for ``template composite dark matter", finding essentially no bound on their model from direct detection. 

\begin{figure}
\label{fig:pol-su4}
\centering
\includegraphics[width=0.6\textwidth]{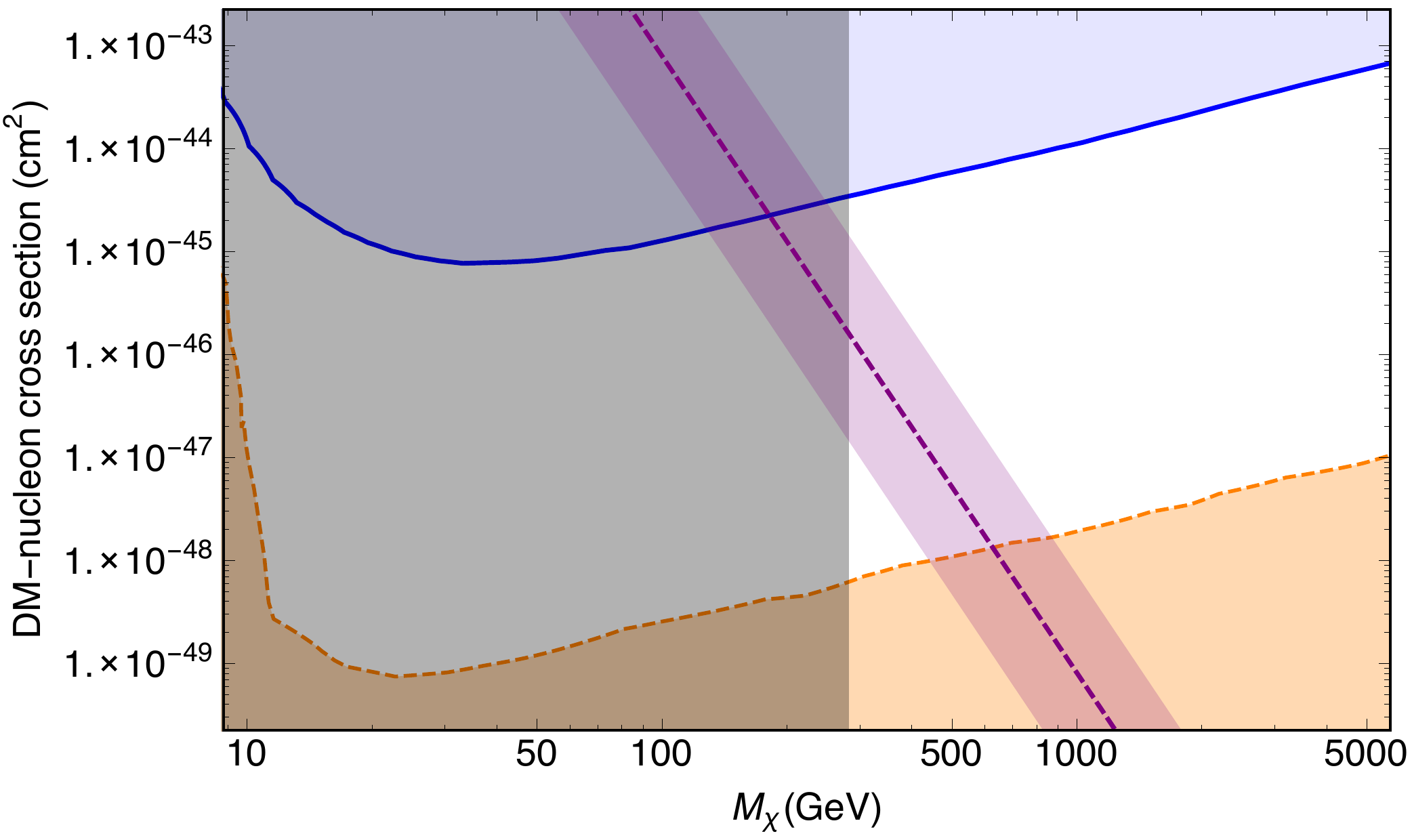}
\caption{Direct detection cross-section prediction for xenon (purple band) for stealth dark matter interacting through electromagnetic polarizability, calculated using lattice results \cite{Appelquist:2015zfa}.  The blue shaded region (top) indicates current experimental bounds from LUX \cite{Akerib:2013tjd}; the grey region (left) shows collider bounds on charged mesons in this model; the orange region (bottom) shows the anticipated irreducible cosmic neutrino background.}
\end{figure}

\subsection{Higgs interaction}

If the composite dark sector contains fundamental fermions $f$, it is natural for them to obtain some of their mass from a Yukawa coupling $y_f$ to the Higgs boson, inducing a mass of order $m_f \sim y_f v$.  If this coupling is present, it will induce a Higgs coupling to any composite state, e.g. a dark baryon $B$ formed from the $f$ fields, of the form
\beq
\sum_f y_f \langle B | \bar{f} f | B \rangle.
\eeq
This mirrors the way in which the Higgs couplings of the proton and neutron in the standard model arise; they depend on the individual quark Yukawa couplings, and on the scalar-current matrix element, also known as the ``sigma term''.

The coupling to the Higgs need not be the only source of mass for the $f$ fermions; they may also have (technically natural) vector-like mass terms, or Yukawa couplings to other new scalar fields.  In general, we can parameterize the fraction of the fermion mass which is due to the Higgs field by defining the parameter\cite{Appelquist:2014jch}
\beq
\alpha = \frac{v}{m_f} \left. \frac{\partial m_f(h)}{\partial v} \right|_{h=v}
\eeq
where $m_f(h) = m + yh/\sqrt{2}$, and $m$ encapsulates other sources of mass.  This parameter varies from $\alpha=0$ if $y=0$ (no Higgs contribution to $m_f$), to $\alpha=1$ when $m=0$ (so the Higgs boson is the only source of mass for $f$.)

If the ratio $m_f / m_B$ is kept fixed, then the direct detection cross section for the dark baryon $B$ increases quadratically with $m_B$, leading to fairly strict bounds from current experiments when $\alpha$ is large.  In particular, comparison with LUX yields the bound \cite{Appelquist:2014jch}
\beq \label{eq:alpha-bound}
\alpha \lesssim \left( \frac{370\ \rm{GeV}}{m_B} \right)^{1/2} \times \begin{cases}
0.34& m_{PS} / m_V = 0.55, \\
0.05& m_{PS} / m_V = 1,\end{cases}
\eeq
where $m_{PS} / m_V$ is the ratio of pseudoscalar to vector meson mass in the SU$(4)$ theory considered, which is a proxy for $m_f / m_B$.  These results strongly disfavor $\alpha = 1$, i.e. a purely electroweak origin for the dark sector fermion masses is essentially ruled out.

\begin{figure}
\label{fig:sigma-lat}
\centering
\includegraphics[width=0.6\textwidth]{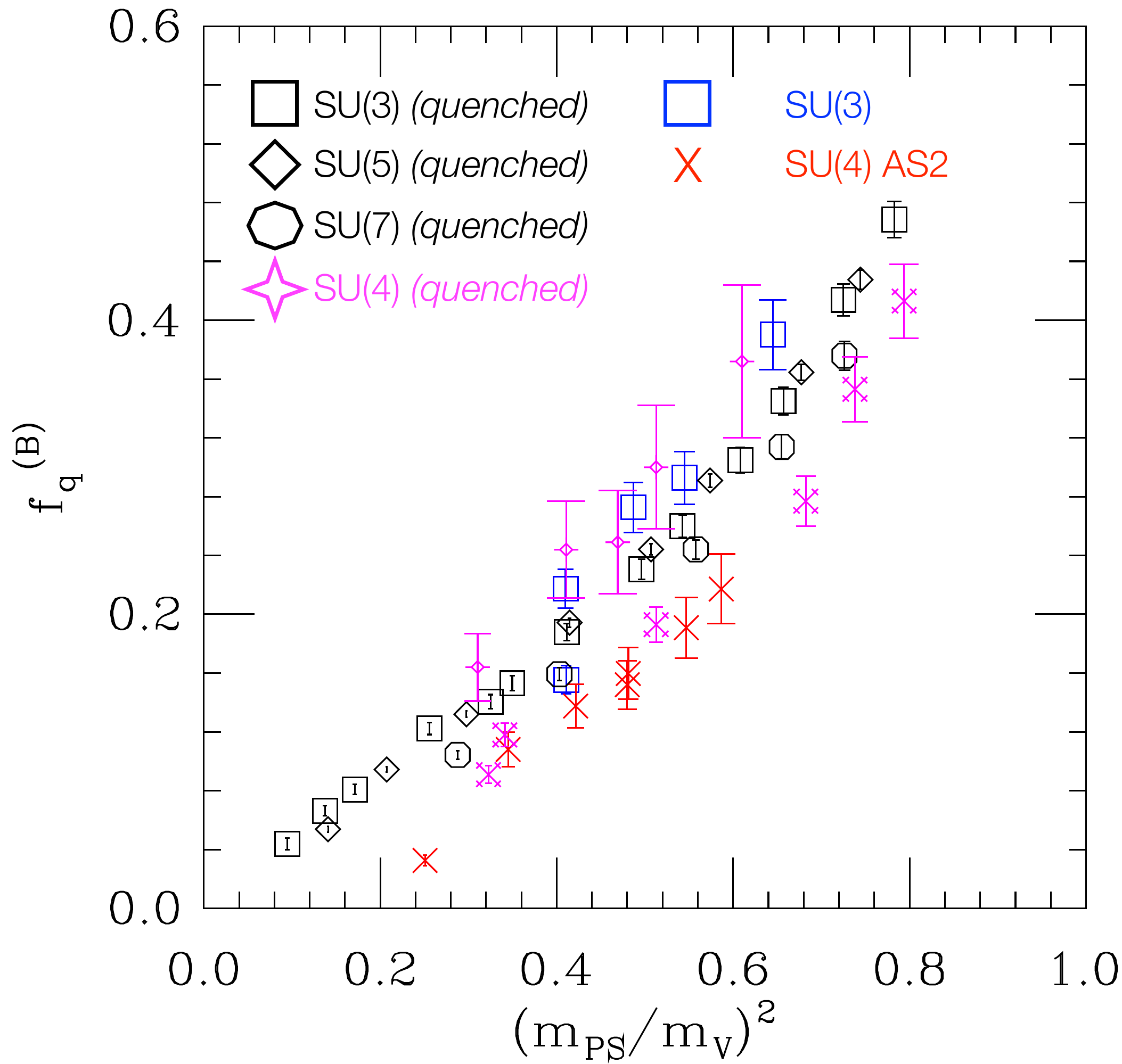}
\caption{Results obtained from lattice simulations\cite{DeGrand:2015lna} in various theories for the ``dark sigma term" $f_f^{(B)}$, defined in the text.  The results are generally quite consistent as a function of fermion mass, even as the gauge group and fermion representation are varied.}
\end{figure}

Although this result assumes a particular dark sector model based on SU$(4)$ gauge theory, there is some evidence that the constraint $\alpha < 1$ is fairly robust.  The main non-perturbative input which gives the bound Eq.~\ref{eq:alpha-bound} is the ``dark sigma term"
\beq
f_f^{(B)} \equiv \frac{\langle B | m_f \bar{f} f | B \rangle}{m_B} = \frac{m_f}{m_B} \frac{\partial m_B}{\partial m_f},
\eeq
applying the Feynman-Hellmann theorem to obtain the last equality.  This quantity is readily determined from lattice spectroscopy of the baryon mass vs. input fermion mass.  Lattice results for a number of different gauge theories \cite{DeGrand:2015lna} are shown in Fig.~\ref{fig:sigma-lat}, and indicate that for similar mass ranges, the non-perturbative value of $f_f^{(B)}$ obtained tends to be consistent across different strongly-coupled theories.

\section{Collider implications}
	\label{sec:DM_other}
	
Collider experimental probes of strongly-coupled composite dark matter
have also been considered in the distinct parametric regimes 
of the theory:  pion-like, quarkonia-like, and intermediate
or mixed cases.  In addition to the dark confinement scale
$\Lambda_d$ and the dark fermion mass(es) $m_q$, there is also
the energy scale of a typical production cross section, 
such as $\sqrt{s}$ of an $e^+ e^-$ collider or the partonic
equivalent, $\sqrt{\hat{s}}$, of a hadron collider.
Qualitatively different phenomena result depending on 
the relative sizes of the dark scales $\Lambda_d$, $m_q$, 
and the characteristic energy scale of the collider.

In addition to direct collider searches for composite dark sector particles,
other indirect constraints can also be obtained from colliders.  
In particular, if the dark sector has electroweak charges, 
electroweak precision observables such as the $S$,$T$,$U$ 
parameters \cite{Peskin:1991sw} can give important constraints.  
The details of this constraint are strongly model-dependent, 
so we do not consider it further here, but we note that given 
a specific composite dark matter model, the $S$-parameter contribution 
can be computed on the lattice from current-current correlation 
functions \cite{Shintani:2008qe,Boyle:2009xi}.

\subsection{Quirky signals}

A dark hidden sector with a new non-Abelian force that confines
can leave very unusual collider phenomenology.  
This was recognized long ago in the context of 
``hidden valley'' models.\cite{Strassler:2006im,Han:2007ae}
A striking example was emphasized in the context
of ``quirky'' models\cite{Kang:2008ea} -- a dark sector contains
dark fermions that transform under part of the standard model, 
while being deep in the quarkonia regime, $\Lambda_d \ll m_q$. 
Here $m_q \lesssim \sqrt{s}$ ($\sqrt{\hat{s}}$) so that 
pairs of dark fermions could be produced easily by
Drell-Yan or other standard model processes.
With a dark confinement scale much smaller than the dark fermion
mass scale, the dark color strings cannot fragment, and so the 
dark fermions can travel macroscopic distances while still held 
together by a very weak unbreakable dark color string.  
Depending on the standard 
model charges of the dark fermions, this can leave highly exotic tracks 
and energy deposition in detectors that is unlike anything 
produced in the standard model.\cite{Kang:2008ea,Harnik:2008ax}
Once the dark fermions are sufficiently heavy, they can be
integrated out, giving effective operators between 
standard model fields and the dark glue fields. 
This can provide an opportunity to probe glueball
phenomenology at colliders \cite{Juknevich:2009ji}.
Additionally, models with $\Lambda_d \ll m_q \simeq \sqrt{s}$ can give
glueball dark matter, so long as $\Lambda_d$ is
sufficiently small that the glueball has a lifetime longer than
the age of the universe.

\subsection{Dark shower signals}

Qualitatively distinct phenomenology can occur in a different regime
(e.g., see\cite{ArkaniHamed:2008qn,Baumgart:2009tn})
in which $m_q \lesssim \Lambda_d \ll \sqrt{s}$ ($\sqrt{\hat{s}}$), 
where the dark non-Abelian sector is expected to shower,
fragment, hadronize, and decay (for the states that have
no conserved quantum
number)\cite{Carloni:2010tw,Carloni:2011kk,Schwaller:2015gea,Cohen:2015toa}.
One possibility follows from production of dark fermions that
subsequently shower in the dark sector, followed by 
decays of dark mesons back to standard model jets.
In the case where the decay lengths of the dark mesons are
macroscopic, this can give emerging jet signals 
at colliders.\cite{Schwaller:2015gea}
Another possibility is that there are stable dark mesons -- 
dark matter -- also produced during the dark shower. 
In this case, the dark matter is produced in an ordinary 
QCD-like parton shower along with other light degrees 
of freedom that decay hadronically.  The result is a 
multijet plus missing transverse momentum signature 
where one of the jets is closely aligned with the direction
of the missing momentum, called a ``semi-visible'' 
jet\cite{Cohen:2015toa}.

\subsection{Meson production and decay}

In the regime where $\Lambda_d \sim m_q \sim \sqrt{s}$ ($\sqrt{\hat{s}}$),
meson production and decay is likely the most promising way to
probe confining, non-Abelian dark sectors.  
At LEP II and the LHC, with $\sqrt{s} \sim \sqrt{\hat{s}} \sim v_{\rm EW}$, 
the signals bear some resemblance to the older studies of technicolor 
theories where the meson phenomenology dominates the experimental 
observables.\cite{Hill:2002ap,Martin:2008cd,Barbieri:2010mn,Andersen:2011yj,Brod:2014loa}
This is not hard to understand - in general collider experiments can 
much more easily produce dark mesons than dark baryons 
(just like their QCD analogues).  In other cases where the dark 
sector mass scales are smaller, for example 
$\Lambda_d \sim m_q \sim 1$~GeV (where dark matter self-interactions
could affect small-scale structure), the spectroscopy of these
theories could be probed by lower energy experiments, 
such as high luminosity $b$-factories\cite{Hochberg:2015vrg}.

Composite dark sectors contain a large number of resonances.
Among the mesons, both the (pseudo)scalar and vector mesons 
provide excellent opportunities for collider studies.
The scalar mesons are generally the lightest new particles 
in theories with $m_q \lesssim \Lambda_d$ as demonstrated 
by lattice simulations.  Vector mesons also provide an
excellent probe of composite dynamics, especially when 
there is some effective kinetic mixing between the photon 
or electroweak gauge bosons and the new vector mesons.

In the following, we consider a study of one concrete example,
the lightest charged meson in Stealth Dark Matter.

\subsubsection{Case Study Example:  Lightest Stealthy Mesons}

In Stealth Dark Matter, the lightest charged meson is a $0^{-+}$
that we denote by $\Pi^{\pm}$.  In contrast to e.g. 
supersymmetric extensions of the standard model where the lightest
supersymmetric particle serves as a dark matter candidate, 
in composite models these charged states can be significantly 
lighter than the dark matter itself.  Direct searches for 
charged states can therefore have better reach than generic 
missing-energy searches for certain composite models.

As a first approximation, treating the $\Pi^\pm$ as point-like scalars 
carrying unit electric charge, the production cross-section 
from electron-positron collisions is
\beq
\sigma(e^+ e^- \rightarrow \Pi^+ \Pi^-) = \frac{\pi \alpha^2}{8E^2} \left(1 - \frac{M_{\Pi}^2}{E^2} \right)^{3/2},
\eeq
where $E$ is half of the center-of-mass energy of the collision, 
roughly 100 GeV for LEP.  This gives e.g. a cross section of 0.2 pb 
with $M_\Pi = 80$ GeV; since the LEP experiments recorded 
approximately 1000 pb${}^{-1}$ of integrated luminosity, 
hundreds of candidate events would have been produced.

If the $\Pi^{\pm}$ are stable on collider timescales, then searches 
for charged tracks can give a bound.  
On the other hand, with appropriate electroweak couplings the 
$\Pi^{\pm}$ can decay to standard model particles via annihilation 
of its constituent fermions into a $W$ boson \cite{Appelquist:2015yfa}. 
Because the initial state is spin-zero, this decay proceeds only 
through the longitudinal part of the $W$, if the decay is into a 
fermion doublet $ff'$, the width will be proportional to the 
final-state fermion masses: with $m_f \gg m_{f'}$,
\beq
\Gamma \propto m_f^2 \left( 1 - \frac{m_f^2}{M_\Pi^2} \right)^2.
\eeq
Again focusing on LEP, for $M_\Pi$ of order 100 GeV, the branching of $\Pi^{\pm}$ decays is roughly 70\% into $\tau \bar{\nu}_\tau$, and 30\% into $c\bar{s}$ pairs.  Combined with the large production cross-section, this leads to a robust constraint from stau searches at LEP \cite{Heister:2001nk,Heister:2003zk,Abdallah:2003xe,Achard:2003ge,Abbiendi:2004gf} that require $M_\pi \gtrsim 90$ GeV (assuming that the decay of $\Pi$ is prompt).  More restrictive bounds may be obtainable by using LHC data, but would require a more detailed study.

The translation of this bound into a bound on the dark matter mass itself depends on the spectrum of the dark sector.  If the composite dark matter candidate is the lightest meson, then the $\Pi^{\pm}$ will tend to be nearly degenerate with it, so that the bound applies directly to $M_{DM}$.  (Exceptions are possible, for example, in models which are embedded directly into electroweak symmetry breaking, the $\Pi^{\pm}$ can become the longitudinal modes of the $W$ and $Z$ bosons \cite{Ryttov:2008xe,Belyaev:2010kp}, removing the bound.)  On the other hand, if the dark matter is baryonic, then it will tend to be heavier than $\Pi^{\pm}$, so that the bound is stronger; for example in \cite{Appelquist:2015yfa}, the dark matter mass bound is $M_{DM} \gtrsim 250-320$ GeV, depending on the specific model parameters which determine $M_{DM} / M_\Pi$.  For the case of glueball dark matter, any charged fermions will be much heavier than the dark matter candidate by necessity to ensure its stability, which again effectively removes the bound.

Going forward, it is important to note that this analysis is relatively simplistic.  In particular, we have treated the $\Pi^{\pm}$ as point-like charged particles, but their interactions with the photon will actually be proportional to a momentum-dependent form factor $F(q^2)$.  The form factor satisfies $F(0) = 1$ - that is, at zero momentum transfer the $\Pi^{\pm}$ do appear point-like - but for e.g. Drell-Yan photoproduction, the momentum transfer at the vertex becomes $q^2 = 4M_\Pi^2$ at threshold, and the form factor may be significantly different.  (In QCD the form factor is larger at this value of $q^2$, growing to $|F| \sim 6$ due to the dominant effects of the $\rho$ vector meson in this channel \cite{Feng:2014gba}.)

Lattice calculation of such form factors would allow for more accurate calculations of the production of the $\Pi^{\pm}$.  However, lattice simulations are carried out in Euclidean spacetime, which means that only spacelike form factors ($q^2 < 0$) are readily accessible.  Lattice studies both for QCD (reviewed in \cite{Brandt:2013ffb}) and for SU$(2)$ gauge theory \cite{Hietanen:2013fya} have shown good agreement with vector-meson dominance (VMD) models in the spacelike region, although far from the vector ($\rho$) resonance itself.  The more challenging direct calculation of the timelike form factor using the L\"{u}scher finite-volume method has been demonstrated recently in lattice QCD \cite{Feng:2014gba}.  A similar calculation in a different strongly-coupled theory could provide direct input for collider studies, as well as giving an interesting test of vector-meson dominance away from QCD.

\section{Outlook}
	\label{sec:DM_outlook}
	There are several interesting research directions which remain largely unexplored, either from the phenomenology or lattice side, or both.  The elastic scattering cross section of composite states is an interesting quantity to compute on the lattice, determining the strength of dark matter self-interactions in the present universe.  The leading scattering phase shifts are readily accessible through the L\"{u}scher finite-volume method, and several lattice QCD calculations of meson-meson and baryon-baryon scattering have now been performed \cite{Detmold:2015jda}.  Application of the same technique to glueball-glueball scattering faces no theoretical obstacles and would also be quite interesting.  Studies of meson-meson scattering in heavier mass regimes, where chiral perturbation theory cannot be used effectively, would be particularly useful in the context of mesonic composite dark matter models.

We have not discussed indirect detection in this review.  If the composite dark matter is a thermal relic, then it can annihilate with its antiparticle, potentially producing an observable astrophysical signal.  For annihilation directly to standard model final states, the relevant process is just the time-reversal of collider production, for example $X^\dagger X \rightarrow e^+ e^-$.  If this occurs primarily through e.g. an $s$-channel photon, then it can be calculated from the time-like electromagnetic form factor of $X$.  As mentioned in the context of collider studies, this form factor has been calculated in the timelike region in lattice QCD recently \cite{Feng:2014gba}.

In general, particularly for baryon-like dark matter, the dominant annihilation channel is likely into lighter composite states, which would then decay into standard model products.  This is a very difficult process to study, particularly since based on large-$N_c$ arguments \cite{Witten:1979kh} and QCD experiments \cite{Amsler:1991cf,Dover:1992vj}, at low energies the strongest annihilation is generally into multiple final-state mesons.   Such a process is extremely difficult to study using current lattice techniques.  However, it would nevertheless be interesting to explore the phenomenological implications of $2 \rightarrow N$ annihilation channels in the context of indirect detection, even in the absence of quantitative predictions for the rate in a given strongly-coupled model.

Study of the finite-temperature properties of general strongly-coupled gauge theories can have interesting implications for physics in the early universe.  One particularly interesting possibility that has been noted is the creation of primordial gravitational waves from such a strong sector if it undergoes a first-order phase transition in the early universe \cite{Schwaller:2015tja,Caprini:2015zlo}.  Lattice calculations can reliably determine the order of the finite temperature transition in a given model, and may furthermore be able to give quantitative information on the predicted gravitational wave spectrum based on thermodynamic quantities such as pressure differences on either side of the transition.

There are several interesting potential connections between lattice simulations and other specific dark matter models or phenomena that we haven't covered in the rest of this review.  One specific example is given by ``neutral naturalness'' models,\cite{Chacko:2005pe,Foot:2014mia,Craig:2015pha,Garcia:2015loa,Garcia:2015toa,Farina:2015uea,Freytsis:2016dgf} in which the twin Higgs sector can contain a twin version of QCD, containing states with potentially interesting collider signals, e.g. \cite{Craig:2015pha,Cheng:2015buv} such as twin glueballs and twin quarkonia.  Lattice QCD results have sometimes been applied to estimate spectra in these theories, but because the masses of the twin quarks relative to the twin confinement scale can be very different from ordinary QCD, there is room for lattice studies in different parametric regimes to make an impact in these theories.

Although it is not a composite dark matter candidate itself, axion dark matter also depends on non-perturbative physics that can be treated using lattice calculations.  In particular, the topological susceptibility of the QCD vacuum at finite temperature determines the thermal mass of the axion, which is a crucial input in determination of the axion relic density.  Recent lattice calculations have begun to pursue this quantity non-perturbatively \cite{Berkowitz:2015aua,Kitano:2015fla,Borsanyi:2015cka,Trunin:2015yda,Bonati:2015vqz}, finding intriguing preliminary results which promise to significantly improve on earlier estimates using purely phenomenological models.

\section*{Acknowledgements}
    \label{sec:acknowledgements}

We thank 
Y.~Bai, 
T.~Cohen, 
Y.~Hochberg, 
R.~Lewis, 
M.~McCullough, 
M.~Pospelov, 
E.~Rinaldi, 
F.~Sannino, and  
T.~Tait 
for providing helpful comments, corrections, and suggestions for 
improvement on a preliminary version of this review.  The authors are
supported in part by the U.~S.~Department of Energy under contract
nos. DE-SC0011640 (GDK) and DE-SC0010005 (ETN).  Brookhaven 
National Laboratory is supported by the DoE under contract no.~DE-SC0012704.

\bibliography{lattice_dm}

\end{document}